\documentclass[final,5pt,twocolumn]{elsarticle}

\usepackage{lineno,hyperref}
\usepackage{hyperref}
\usepackage[shortcuts]{extdash}
\usepackage{amsmath}
\journal{Journal of Alloys and Compounds}

\usepackage[margin=1.5cm]{geometry}%2.5
\usepackage{graphicx}
\usepackage{multicol}
\usepackage{subfigure}

\usepackage{color}
\usepackage[usenames, dvipsnames, svgnames, table]{xcolor}

\makeatletter
\newenvironment{tablehere}
  {\def\@captype{table}}
  {}

\newenvironment{figurehere}
  {\def\@captype{figure}}
  {}
\makeatother
\begin{document}

\begin{frontmatter}

\title{Kondo Lattice Behavior Observed in the CeCu$_9$In$_2$ Compound}


%\author[uj]{R.~Kurleto\corref{mycorrespondingauthor1}}
\author[uj]{R.~Kurleto}
\ead{rafal.kurleto@uj.edu.pl}
\author[uj]{A.~Szytu\l{}a}
\author[us]{J.~Goraus}
\author[uj]{S.~Baran}
\author[ul]{Yu.~Tyvanchuk}
\author[ul]{Ya.~M.~Kalychak}
\author[uj]{P.~Starowicz}
\ead{pawel.starowicz@uj.edu.pl}

\address[uj]{Marian Smoluchowski Institute of Physics, Jagiellonian University, {\L}ojasiewicza 11, 30-348 Krak{\'o}w, Poland}
\address[us]{Institute of Physics, University of Silesia, 75 Pu{\l}ku Piechoty 1a, 41-500 Chorz{\'o}w, Poland}
\address[ul]{Department of Analytical Chemistry, Ivan Franko National University of Lviv, Kyryla and Mephodiya 6, 79005 Lviv, Ukraine }

\begin{abstract}
We report systematic studies of CeCu$_9$In$_2$, which appears to be a new Kondo lattice system. Electrical resistivity exhibits a logarithmic law characteristic of Kondo systems with a broad maximum at $T_{coh}\approx$45~K and it obeys the Fermi liquid theory at low temperature. Specific heat of CeCu$_9$In$_2$ is well described by the Einstein and Debye models with electronic part at high temperature. Fitting of the Schottky formula to low temperature 4f~contribution to specific heat yielded crystal field splitting of 50.2~K between a doublet and quasi-quartet. The Schotte-Schotte model estimates roughly Kondo temperature as $T_K\approx$5~K, but does not reproduce well the data due to a sharp peak at~1.6~K. This structure should be attributed to a phase transition, a nature of which is possibly antiferromagnetic. Specific heat is characterized with increased Sommerfeld coefficient estimated as $\gamma\approx$132~mJ/(mole$\cdot$K$^2$). Spectra of the valence band, which have been collected with ultraviolet photoelectron spectroscopy (UPS), show a peak at binding energy$\approx$250~meV, which originates from the Ce~4f electrons and is related to the 4f$^1$$_{7/2}$ final state. Extracted 4f~contribution to the spectral function exhibits also the enhancement of intensity in the vicinity of the Fermi level. Satellite structure of the Ce~3d levels spectra measured by X-ray photoelectron spectroscopy (XPS) has been analyzed within the framework of the Gunnarsson-Sch{\"o}nhammer theory. Theoretical calculations based on density functional theory (FPLO method with LDA+U approach) delivered densities of states, band structures and Fermi surfaces for CeCu$_9$In$_2$ and LaCu$_9$In$_2$. The results indicate that Fermi surface nesting takes place in CeCu$_9$In$_2$.
\end{abstract}


\end{frontmatter}

%\linenumbers

\section{Introduction}
	Presence of unpaired 4f electron in cerium atom leads to many unusual physical phenomena in cerium intermetallic compounds. Often, experimental evidence points to a dichotomous character of 4f electrons, which are neither fully localized nor fully itinerant. This plays an important role in physics of such systems. These electrons form a relatively narrow band, which hybridizes with carriers from a conduction band. This coupling can lead to quenching of magnetic moments of~f~electrons and to formation of singlets in a ground state. Other complex states, such as a Kondo lattice or a mixed valency are also possible. The aforementioned dynamic singlet state in a Kondo regime is manifested by a pronounced anomaly in an electrical resistivity and by the so called Kondo peak in a spectral function. While the presence of the latter feature in a single impurity regime (Kondo regime) is well established, analogous spectral shape in the case of a Kondo lattice may be an interesting subject of research.

	Crystal chemistry of ternary indides with rare-earth and transition metal elements was the subject of previous studies due to intriguing physical properties~\cite{kaliczak2005, bigun2013, bigun2014}. The structural phase diagram of the Ce--Cu--In system was constructed previously and extensively discussed~\cite{kaliczak2005, kaliczak91, kaliczak98}. CeCu$_9$In$_2$ compound crystallizes in the YNi$_9$In$_2$-type tetragonal structure (P4/mbm space group), with lattice constants: a=8.5403(4)~{\AA} and~c=5.0204(5)~\AA. LaCu$_9$In$_2$ is isostructural to CeCu$_9$In$_2$. Its crystal structure, specific heat and magnetic susceptibility were reported before~\cite{baran2016}. Lattice constants of LaCu$_9$In$_2$ slightly differ from those obtained for the cerium counterpart (a=8.6351(4)~\AA,~c=5.1488(3)~\AA). Both intermetallic compounds are characterized by narrow homogeneity ranges with substitution of Cu by In~\cite{kaliczak2005}. Specific heat of LaCu$_9$In$_2$ is well modeled by the sum of phononic and electronic contributions (details of the fitting procedure are available elsewhere~\cite{baran2016}). The magnetic susceptibility of LaCu$_9$In$_2$, measured at~0.1~T and~1~T, is almost temperature independent. This fact was interpreted as a symptom of a Pauli paramagnetism~\cite{baran2016}. It should be mentioned that replacement of a transition metal may have a significant effect; recently, a~mixed valence state has been proposed for the isostructural CeNi$_9$In$_2$ compound~\cite{bigun2014, moze95,szytula2014, kurleto2015}.

	Magnetic properties of CeCu$_9$In$_2$ were studied before~\cite{baran2016}. Measured magnetic susceptibility follows the Curie-Weiss law in a broad temperature range~(100-350~K). The estimated Curie-Weiss temperature is equal to~$-39$~K, while inferred value of effective magnetic moment is slightly enhanced to~$\mu$=2.88~$\mu_B$, which can be compared to $\mu$=2.54~$\mu_B$, what is a theoretical value for free~Ce$^{3+}$~\cite{baran2016}. This value suggests that valency of cerium is close to~3.
	
	In this article we describe physical properties of the CeCu$_9$In$_2$ compound. Namely, we provide electrical resistivity, specific heat, as well as photoelectron spectra of the valence band and Ce~3d levels. In order to reveal the role of the 4f~electrons we have also performed measurements on isostructural LaCu$_9$In$_2$. The obtained spectra are confronted with the results of theoretical calculations. We have analyzed our data in terms of the Kondo lattice with the Fermi liquid ground state. Our results testify significant coupling between the 4f electrons and carriers from the conduction band in CeCu$_9$In$_2$. This compound is a new system in which a realization of a Kondo lattice takes place.

\section{Materials and methods}
	Fabrication and characterization of polycrystalline samples of CeCu$_9$In$_2$ and LaCu$_9$In$_2$ (the latter with exact stoichmiotery: LaCu$_{8.25}$In$_{2.75}$) were described elsewhere~\cite{baran2016}. Electrical resistivity has been measured with application of four-terminal alternating current technique in Physical Property Measurement System (PPMS, Quantum Design) in temperature range 2--300~K. Specific heat has also been measured in PPMS, with application of a relaxation method~(two-tau model), in similar temperature range. Additional measurement of specific heat in temperature range 0.4--10~K has been performed  with application of $^3$He refrigerator.

	Photoelectron spectroscopy has been conducted with an in-house experimental setup, equipped with VG~Scienta R4000 photoelectron energy analyzer. X-ray photoelectron spectroscopy (XPS) has been performed with application of  Mg~K$_{\alpha}$ (h$\nu$=1253.6~eV) and Al~K$_{\alpha}$ (h$\nu$=1486.6~eV) radiation (without a monochromator) at temperature: 12.5~K, 100~K or 292~K. He~I (h$\nu$=21.2~eV) and He~II (h$\nu$=40.8~eV) spectral lines from helium lamp have been used in the ultraviolet photoelectron spectroscopy~(UPS) studies. Spectra have been collected at the same temperatures as the ones from XPS.  Both XPS and UPS measurements do not reveal any significant change in the function of temperature, so further in the article we present only spectra collected at the lowest possible temperature~(12.5~K). Base pressure during measurements was equal to 5$\cdot$10$^{-11}$~mbar. Before measurements, surface of the samples was polished under ultra high vacuum conditions (base pressure~2$\cdot$10$^{-10}$~mbar) with application of a diamond file. Calibration was provided by a measurement of binding energy of the Au~4f states on polycrystalline gold layer and by measurement of Fermi edge of polycrystalline Cu for XPS and UPS measurements, respectively.

	Full potential local orbital (FPLO) code~\cite{koepernik99} in scalar relativistic version has been used in order to obtain partial densities of states (DOS) of CeCu$_9$In$_2$ and LaCu$_9$In$_2$. Additional correlations have been involved in around mean field scheme with application of local spin density approximation (LSDA+U)~\cite{anisimov93}. Perdew-Wang exchange-correlation potential~\cite{perdew92, ceperley80} has been assumed in calculations. In case of the compound with cerium, DOS have been obtained for Coulomb repulsion on the f~shell equal to~0~eV,~2~eV (not shown in this paper) and~6~eV. The band dispersion along important crystallographic directions in k-space as well as Fermi surface (FS) of both studied compounds have been calculated with application of Elk software~\cite{elk}. The results from FPLO and Elk are consistent. All theoretical calculations have been performed assuming non-magnetic state of CeCu$_9$In$_2$, what corresponds to a paramagnetic state for which the photoemission spectroscopy has been realized.

\section{Results and Discussion}
\subsection{Electrical resistivity}
Electrical resistivity of the CeCu$_9$In$_2$ compound, measured as a function of temperature, exhibits a pronounced anomaly~(\hyperref[fig:res]{Fig.~\ref*{fig:res}~a}). There is a broad maximum clearly visible at temperature equal to 45~K. 
At low temperature (T$<$13~K) resistivity is well described by the quadratic function:
\begin{equation}\rho=\rho_0+A T^2.\label{eq:FL}\end{equation}
Obtained residual resistivity~$\rho_0$ is equal to~70.10(1)~~$\mu\Omega\cdot$cm, while~$A$~coefficient reaches~10.2(2)$\cdot 10^{-3}$~$\mu\Omega\cdot$cm$/$K$^2$. This behavior implies that a coherent Fermi liquid is developed below 13~K in the studied system. Temperature variation of electrical resistivity for CeCu$_9$In$_2$ is typical of Kondo lattice systems and the broad maximum yields a coherence temperature of~T$_{coh}$$\approx$45~K. It should be mentioned that the equation~(\ref{eq:FL}) with an additional $T^5$ term representing phonon scattering was also fitted to experimental data but the quality of the fit did not improve considerably.

	We have also measured electrical resistivity of the LaCu$_9$In$_2$ compound (\hyperref[fig:res]{Fig.~\ref*{fig:res}~b}), which has got the same crystal structure as CeCu$_9$In$_2$. Temperature dependence of electrical resistivity for LaCu$_9$In$_2$ does not exhibit any pronounced anomaly. Its shape is typical of a simple metal, besides some small concavity, which is most probably due to s-d scattering. We have found that experimental data are described well by the Grüneisen--Bloch--Mott model~\cite{Mott58,Grimvall81}. The fitted function has got the following form:
\begin{equation}\rho=\rho_0+c\frac{T}{\theta} \int\limits_0^{\frac{\theta}{T}} \frac{x^5 dx}{(e^x-1)(1-e^{-x})}-KT^3,\label{eq:GBM}\end{equation}
where: $\rho_0$ is a residual resistivity, $\theta$ denotes the Debye temperature, $c$ is a parameter related to the electron-phonon  coupling, while $K$ is the Mott coefficient, which describes the strength of interband s-d scattering. The obtained values of the parameters are as follows: $\rho_0$=45.33(2)~$\mu\Omega\cdot$cm, $c$=82.2(8)~$\mu\Omega\cdot$cm, $\theta$=189(2)~K, $K$=1.9$\cdot$10$^{-7}$~$\mu\Omega\cdot$K$^{-3}\cdot$cm. 
	
	The measured electrical resistivities of both compounds have been used in order to estimate magnetic contribution of Ce~4f states ($\rho_m$) to electrical transport in CeCu$_9$In$_2$. It was obtained by a subtraction of LaCu$_9$In$_2$ data from those measured for Ce~counterpart~(\hyperref[fig:res]{Fig.~\ref*{fig:res}~c}). The obtained shape is consistent with previous studies of Kondo lattice systems~\cite{szl}. Beyond the aforementioned broad maximum we have found that logarithmic dependency is obeyed between~58~K and~165~K. The fit of the formula proposed~by~J.~Kondo:
\begin{equation}\rho_m=b_1+ b_2\ln T,\label{eq:kondo}\end{equation}
yielded $b_1$=91.1(3)~$\mu\Omega\cdot$cm and $b_2$=-15.54(5)$~\mu\Omega\cdot$cm.

%fig1.a
\begin{figurehere}
\centering\includegraphics[width=0.8\linewidth]{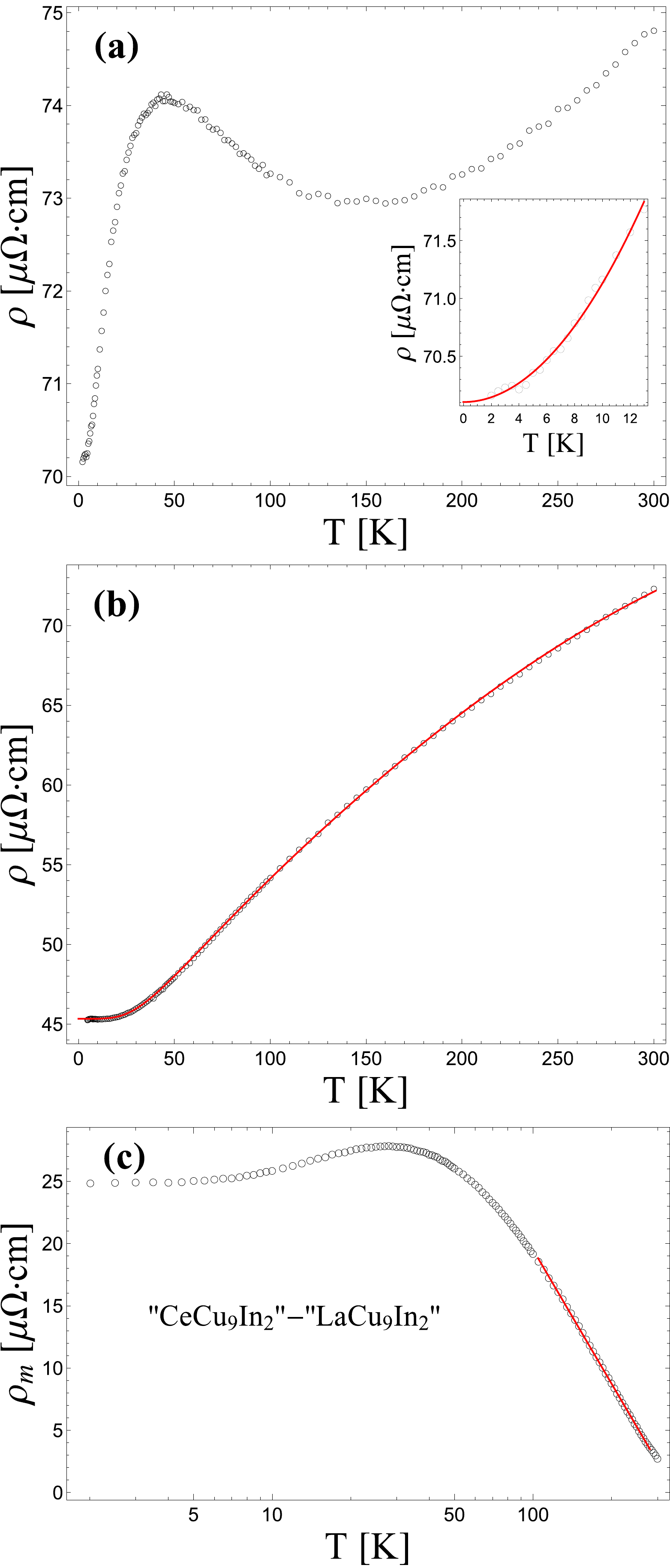}
\caption{(Color on-line)~(a)~Temperature variation of the electrical resistivity (black circles) of CeCu$_9$In$_2$. The red line represents fitted model. Inset: the electrical resistivity at low temperature (2-13~K) with fitted model characteristic of the Fermi liquid~(\ref{eq:FL}). (b)~Temperature variation of the electrical resistivity (black circles) of LaCu$_9$In$_2$. The red line represents fitted  Grüneisen--Bloch--Mott formula~(\ref{eq:GBM}). (c)~Extracted Ce~4f contribution to the electrical resistivity (black circles) of CeCu$_9$In$_2$ with fitted Kondo law~(\ref{eq:kondo}) in temperature range 58-165~K.}
\label{fig:res}
\end{figurehere}

\subsection{Specific heat}
The investigated compounds were also subjected to studies of specific heat (\hyperref[fig:heat]{Fig.~\ref*{fig:heat}}). High temperature part of specific heat (\hyperref[fig:heat]{Fig.~\ref*{fig:heat}~a}) for the CeCu$_9$In$_2$ compound is described well by the sum of lattice contributions (Debye and Einstein models) and electronic part. Namely, we have used the formula:
\begin{equation}c= \gamma' T+9kNR\left(\frac{T}{\theta}\right)^3 \int\limits_0^{\frac{\theta}{T}} \frac{x^4dx}{(e^x-1)^2}+3(1-k)R\left(\frac{\theta_E}{T}\right)^2 \frac{e^{\frac{\theta_E}{T}}}{\left(e^{\frac{\theta_E}{T}}-1\right)^2},\label{eq:DE}\end{equation}
where: $\gamma'$ represents a linear term, $k$ is relative weight of Debye and Einstein modes, $\theta_E$ is the Einstein temperature, $N$=12 is the number of atoms in the unit cell, $R$ is the universal gas constant. Obtained values are as follows: $\theta$=246(8)~K, $\gamma'$=121(8)~mJ/(mole$\cdot$K$^2$), $\theta_E$=176(129)~K, $k$=0.64(3).

At low temperature, we have plotted c/T versus T$^2$ (see inset to \hyperref[fig:heat]{Fig.~\ref*{fig:heat}~a}). We have found a linear dependency in the temperature range between 10~K and 20~K. The linear fit yielded $\gamma=$132(4)~mJ/(mol$\cdot$K$^2$), as well as the Debye temperature equal~to~107.8~K, which is much lower than the value obtained for high temperature fit to the equation~(\ref{eq:DE}). The value of~$\gamma$ is close to $\gamma'$ resulting from fitting more complex formula~(\ref{eq:DE}). Although, the temperatures taken for the linear fit were relatively high for the obtained value of $\theta$~\cite{tran} the fitted $\gamma$ is some estimation of Sommerfeld coefficient. Its high value indicates that charge carriers in CeCu$_9$In$_2$ have got high effective masses. This fact also implies high value of the density of states at the Fermi level. 

Significant anomalies are found at low temperature specific heat. Magnetic contribution related to 4f electrons has been estimated (\hyperref[fig:heat]{Fig.~\ref*{fig:heat}~b}) by a subtraction of the specific heat measured for the isostructural LaCu$_9$In$_2$ compound. One can see three peak-like structures at low temperature. The first one is sharp and located at about~1.6~K (cf.~\hyperref[fig:heat]{Fig.~\ref*{fig:heat}~b}). The contribution from the Kondo effect is expected at such temperature. It can be described in terms of the Schotte-Schotte model~\cite{schotte75}. It was argued before that such model applies even in a Kondo lattice regime: Ce$_{1-x}$La$_x$Ni$_2$Ge$_2$~\cite{pikul2012}. The second, broad peak is located roughly at 22~K. According to our knowledge, it is related to crystal field effects, described by the modified Schottky formula~\cite{souza2016}. The third, additional $\lambda$-type peak at about 7~K is related to the presence of small amount of cerium sesquioxide (Ce$_2$O$_3$)~\cite{huntelaar2000}. It is known, that cerium ion, confined in symmetry different than cubic, should have crystal field configuration, which consists of three doublets. In CeCu$_9$In$_2$, Ce atoms occupy $2a$ site of P4/mbm space group, which is characterized by tetragonal symmetry (site symmetry: 4/m..). However, our data can be described properly by the Schottky formula with a doublet-quartet configuration, which should be observed in case of cubic symmetry. Our attempts of fitting the model with three doublets have led to the results with unreliable coefficients. This discrepancy may be explained by the fact that splitting energy for two doublets is too small to be resolved and therefore a quasi-quartet results from fitting. Similarly, such a quasi-quartet state was observed in case of hexagonal CeCu$_4$Al~\cite{tolinski2009}. Finally, the fitted model has got a form:

\begin{equation}c_{4f}=c_{cf}+c_K+\gamma^{\star} T,\label{eq:c4f}\end{equation}

where $c_{cf}$ stands for contribution related to crystal field:

\begin{equation}c_{cf}=\frac{ 8 e^{-\frac{\Delta}{T}} R \Delta^2}{(4+2e^{-\frac{\Delta}{T}} )2 T^2}\end{equation}

($R$-universal gas constant, $\Delta$-quartet-doublet splitting), while $c_K$ denotes a contribution related to a Kondo scattering given by the Schotte-Schotte model:

\begin{equation}c_K=R\frac{T_K}{T\pi}\left[1-\frac{T_K}{2T\pi}\psi'\left(\frac{1}{2}+\frac{T_K}{2T\pi}\right)\right]\end{equation}

($R$-universal gas constant, $T_K$-Kondo temperature, $\psi'$ – first derivative of digamma function).

%fig2.a
\begin{figurehere}
\centering\includegraphics[width=0.86\linewidth]{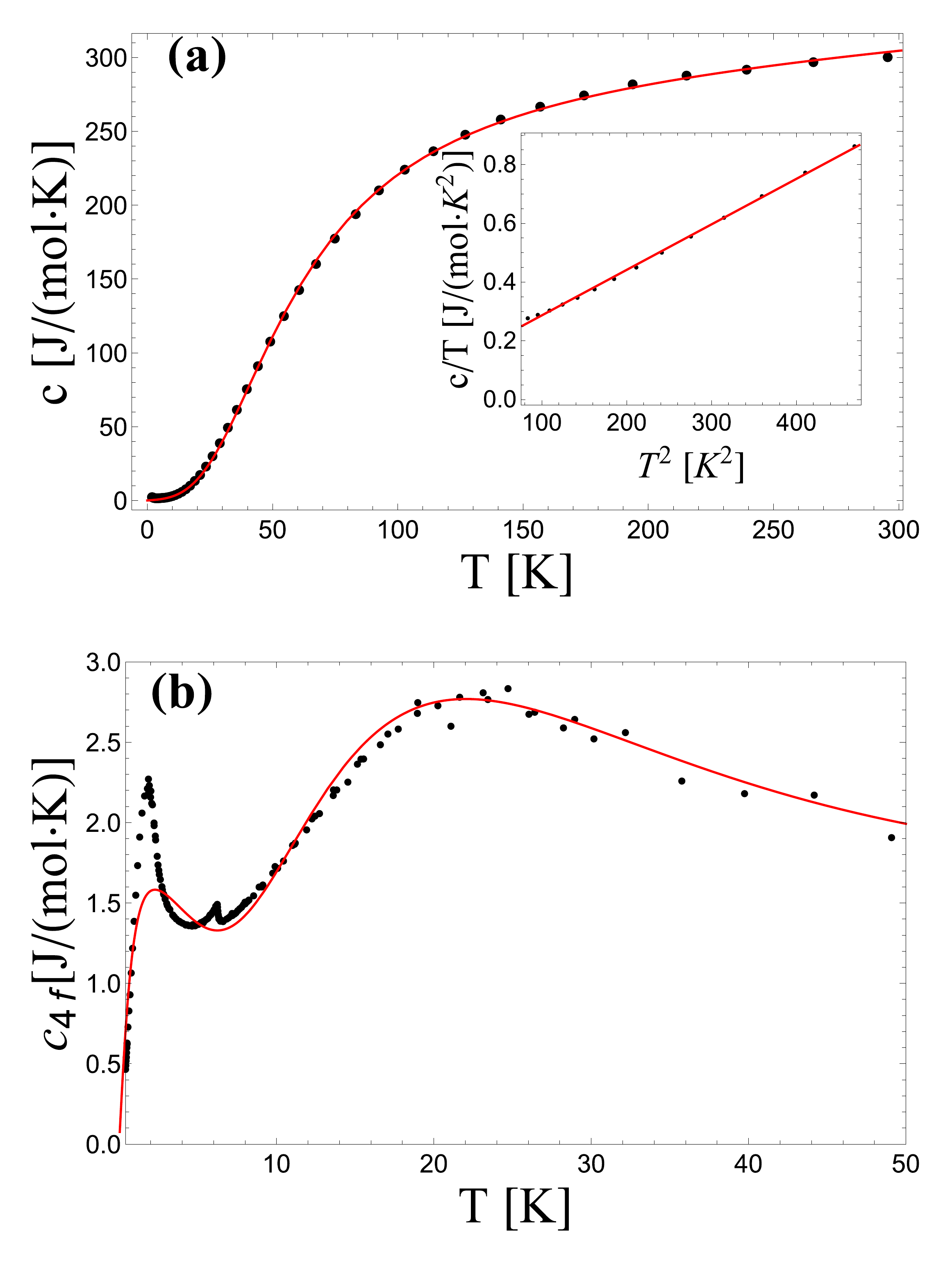}
\caption{(Color on-line) (a)~Specific heat of CeCu$_9$In$_2$ (black points). Red line is a result of the fitted model~(\ref{eq:DE}). Inset: c/T~vs~T$^2$ dependency with the fitted line. (b)~Magnetic contribution of 4f electrons to the specific heat of CeCu$_9$In$_2$ (black points). The red line corresponds to the fit of the sum of contributions from the Shotte-Schotte model and the Schottky model~(\ref{eq:c4f}).}
\label{fig:heat}
\end{figurehere}

%The $a$ coefficient in equation (\ref{eq:c4f}) ensures for the correction for impurities, its value should be close to 1 - we have obtained $a$=1.04(2).
$\gamma^{\star}$ coefficient obtained from the fit of equation ~(\ref{eq:c4f}) to the data equals~13.4(1.9)~mJ/(mole$\cdot$K$^2$). Fitting procedure yielded Kondo temperature~($T_K$) equal to~4.6(3)~K and a doublet-quartet splitting~$\Delta$ equal to~50.2(1.1)~K.  The effect of correlation between charge carriers is included in both Schotte-Schotte and linear contribution to specific heat. Hence, the value of $\gamma^{\star}$ is not a good measure of the mass enhancement. Indeed for $T<<T_K$ the Schotte-Schotte model can be linearized and its contribution to a linear term is not included in $\gamma^{\star}$. Therefore the obtained small value of $\gamma^{\star}$ underestimates the renormalization factor for the electron mass.
	
	One can see that the agreement between the fitted model (eq.~\ref{eq:c4f}, cf.~\hyperref[fig:heat]{Fig.~\ref*{fig:heat}~b}) and experimental data is not perfect, especially below 6~K. In particular, a phase transition, which has been not considered so far, can be reflected in a peak at~1.6~K. In fact, there are premises, which point to the appearance of the transition to antiferromagnetic state in the~CeCu$_9$In$_2$ compound at low temperature. Namely, these are a negative value of the paramagnetic Curie-Weiss temperature and a significant deviation of magnetic susceptibility from the Curie-Weiss law at low temperature~\cite{baran2016}. Magnetic entropy was calculated using estimated~Ce~4f~contribution to the specific heat. At the temperature of the peak at~$T=$1.6~K it reaches about~45\% of a theoretical value~$R$$\ln{2}$, corresponding to a doublet ground state. The reduced value of the magnetic entropy points to significance of the Kondo interaction in the studied compound.

\subsection{Valence band photoelectron spectra}
	Many interesting properties of systems with interacting electrons are reflected in the spectral function. In order to study impact of the Kondo effect on the spectral function we have measured UPS spectra of the valence band of the studied compound. One of the fingerprints of the Kondo effect is the Kondo peak. In heavy fermion compounds maximum of the Kondo peak should lie at $\delta$ – energy above Fermi level according to the Friedel formula~\cite{georges2016}:
\begin{equation}\delta=k_B T_K \sin(\pi n_f),\label{eq:friedel}\end{equation}
where: $n_f$~is a mean occupation of 4f~level, $T_K$~is a Kondo temperature. Usually in photoemission experiments, one can observe only the tail of the Kondo peak, i.e.~the peak is sharply cutted by the Fermi edge. The Kondo peak itself is assigned to the~4f$^1$$_{5/2}$ final state in the UPS spectra of the valence band. For the parameters corresponding to our study; $n_f$=0.96, $T_K$=4.6~K, according to~(\ref{eq:friedel}) $\delta$ is of order of 0.05~meV, which should locate the Kondo peak very close to the Fermi energy.
 
	UPS spectra of the valence band  of CeCu$_9$In$_2$ in vicinity of the Fermi level, measured with He-II and He-I radiation are shown in the \hyperref[fig:UPS]{Fig.~\ref*{fig:UPS}~a}.  In both, He-II and He-I spectra, the tail of the Kondo peak at the Fermi level is not visible. However, in both spectra, one can see some small peak superimposed on the linear slope at binding energy equal to~$-0.28$~eV. This structure can be assigned to 4f$^1$$_{7/2}$ final state in the photoemission process. We  have estimated 4f contribution to the spectral function of CeCu$_9$In$_2$ by subtracting He-I from the He-II spectrum~\cite{yeh93, yeh85}. Such 4f spectral weight yields slightly increased intensities near the Fermi energy (E$_F$) and at~$-0.28$~eV, which are attributed to the~4f$^1$$_{5/2}$ and 4f$^1$$_{7/2}$ final states, respectively. The intensity near~E$_F$ may originate from a Kondo peak, which is hardly visible in this case. 

	UPS spectra of the valence band of isostructural LaCu$_9$In$_2$ have also been measured. One can see that He-II spectrum of LaCu$_9$In$_2$, in contrary to that for CeCu$_9$In$_2$, is completely featureless in the vicinity of  the Fermi level. 4f contribution to the spectral function of CeCu$_9$In$_2$ has also been estimated by subtracting He-II spectrum for LaCu$_9$In$_2$ from He-II spectra for CeCu$_9$In$_2$ (\hyperref[fig:UPS]{Fig.~\ref*{fig:UPS}~b}). The obtained result is similar to the difference between He-II and He-I spectra for CeCu$_9$In$_2$. However, in this case the increased intensity near E$_F$ is not well visible. A sharp Kondo peak is not found. Except for the absence of a sharp, coherent peak at the Fermi level, our spectral function resembles that usually observed for Kondo lattice systems~\cite{starowicz2014}. 
 
%fig3
\begin{figurehere}
\centering\includegraphics[width=0.85\linewidth]{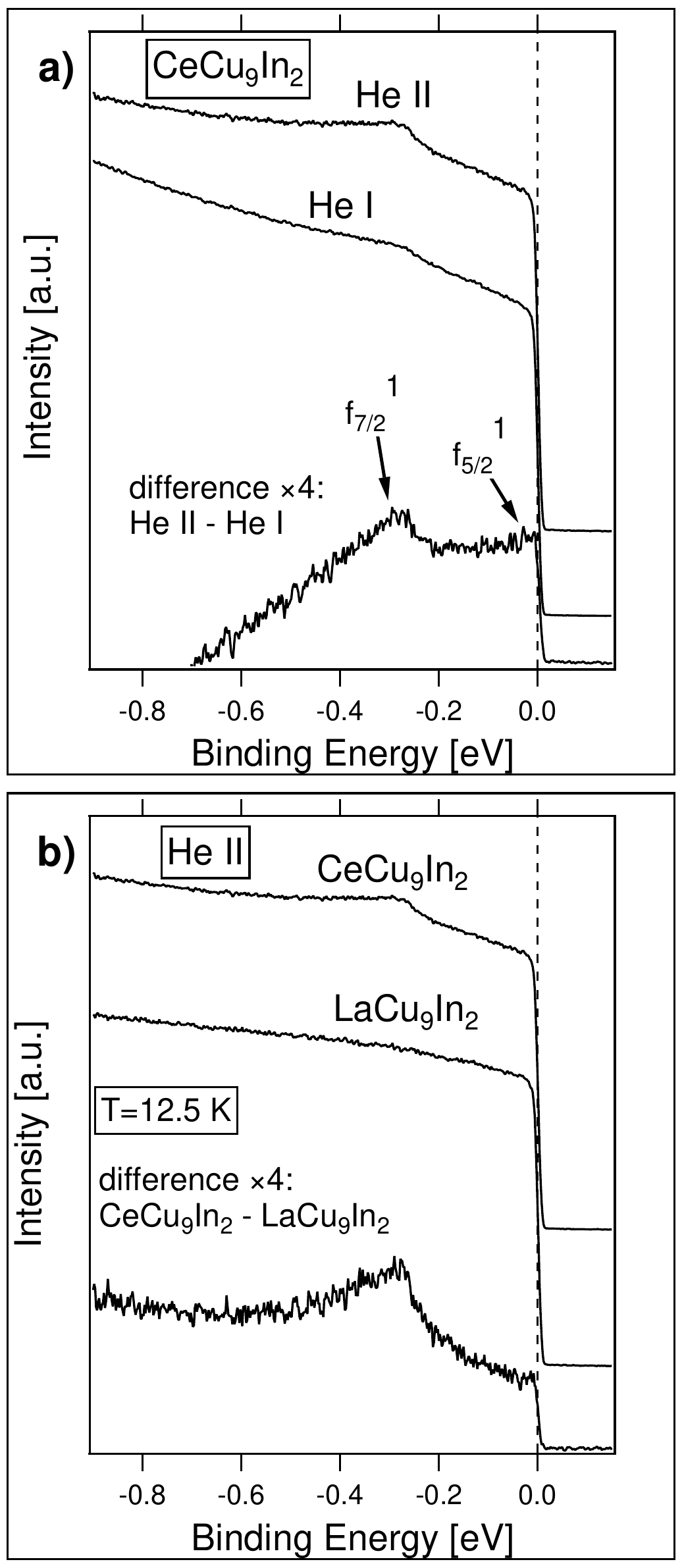}
\caption{(a)~UPS spectra of valence band of CeCu$_9$In$_2$ obtained with He-II (40.8~eV) and He-I (21.2~eV) radiation at the temperature equal to~12.5~K. 4f contribution to the spectral function, estimated by subtraction of He-I spectrum from He-II spectrum, is presented. Peaks corresponding to the f$^1$$_{7/2}$ and f$^1$$_{5/2}$ final states are marked by arrows. (b)~Comparison between CeCu$_9$In$_2$ and LaCu$_9$In$_2$ He-II spectra of valence band. The calculated difference is also an estimation of Ce~4f electrons contribution.}
\label{fig:UPS}
\end{figurehere} 
 
	There can be several reasons which render difficulty in observation of a Kondo peak. For instance, the applied experimental resolution of 13~meV might be not sufficient to record such a sharp feature. Moreover, polycrystalline samples obtained by arc-melting may exhibit certain disorder in particular at the surface, which can supress spectral intensity at~E$_F$. It is also worth to compare spectra presented here for CeCu$_9$In$_2$, with those reported previously for CeNi$_9$In$_2$~\cite{kurleto2015}. In case of the compound with Ni, the Ni~3d states are located almost at the Fermi level and interact strongly with f states resulting in the mixed valence state. In case of compounds with Cu, the~Cu~3d states are rather distant from the Fermi level, making interaction with 4f~states weaker and resulting in the Kondo lattice state. 
	
	We would like to refer to the results reported previously for CeCu$_6$. This system is an archetypal heavy fermion compound, which displays the Kondo lattice behavior without magnetic ordering down to 5~mK~\cite{stewart84, onuki85}. Specific heat as well as magnetic susceptibility of this compound are enhanced at low temperatures. Its electrical resistivity shows a maximum at low temperature, similarly to our results obtained for CeCu$_9$In$_2$. Although fingerprints of Kondo lattice state are clearly visible in electronic transport, the influence of the Kondo effect on the UPS spectra of the valence band of CeCu$_6$ is barely seen in certain reports~\cite{patthey86, patthey90}. Namely, in the He-II spectra, there is no any sharp peak at the Fermi level. However, there is a peak visible at the binding energy equal to~$-0.25$~eV, which was attributed to f$^1$$_{7/2}$ final state. According to the results of inelastic neutron scattering experiment~\cite{goremyczkin93}, the CeCu$_6$ compound has got $T_K=$~5~K. It is commonly believed, that the value of $T_K$ is related to the shape of the UPS spectrum. Namely, the ratio of intensities $I(f^1_{5/2})/I(f^1_{7/2})$ is a monotonic, increasing function of $T_K$. So, when the $T_K$ is small, the intensity of the Kondo peak with respect to its spin-orbit partner is very low. Some studies of heavy fermion systems with low value of $T_K$, state that in order to observe the Kondo peak in the UPS spectra, one should measure with resolution less than $k_BT_K$~\cite{garnier97}. This statement was confirmed by recent measurements on CeCu$_6$, in which the authors reported the Kondo peak in the spectra~\cite{ehm2003, ehm2007}. Possibly, the situation in case of CeCu$_9$In$_2$ is the same as for CeCu$_6$. Analysis of the specific heat of CeCu$_9$In$_2$ yields very low value of $T_K$ (about 5~K), and the He-II spectrum resembles the early spectra of CeCu$_6$.

\subsection{XPS of Ce 3d levels}
	Strong Coulomb interaction between 4f electrons and photoholes created in core levels during photoemission process leads to characteristic satellite structure of the XPS spectrum of Ce 3d levels. Each spin orbit partner in 3d doublet splits into three satellites corresponding to different numbers of electrons on the f shell in the final state of photoemission process. Fitting procedure applied together with the so called Gunnarsson-Sch{\"o}nhammer theory~\cite{fuggle83, gunnarsson83} allows to estimate parameters such as a mean occupation of 4f shell, as well as the strength of the coupling between conduction band and 4f level. 
	
	XPS spectrum of Ce~3d levels of CeCu$_9$In$_2$ is shown in the \hyperref[fig:XPS]{Fig.~\ref*{fig:XPS}}. Peak deconvolution and numerical analysis can be based on the Doniach-\v{S}unji{\'c} theory~\cite{doniach70}. The background has been simulated with application of the Shirley method~\cite{shirley}. Peaks corresponding to the Ce~3d$_{5/2}$ and Ce~3d$_{3/2}$ levels are clearly visible. They are separated by spin-orbit splitting equal to~18.5~eV. Each component of the doublet is further splitted due to the hybridization effects. Namely, each component of the doublet consists of three satellite lines: f$^0$, f$^1$ and f$^2$. The~3d$^9$4f$^0$ component is an evidence of the intermediate valence state of cerium ions, while the~3d$^9$4f$^2$ component gives information about the hybridization between 4f states and conduction electron states. Locations as well as intensities of particular peaks are collated in \hyperref[table:xps]{Table~\ref*{table:xps}}.
	
%xps
\begin{tablehere}
\centering
\caption{Binding energies ($E_B$) and relative intensities ($I$) of satellite lines in XPS spectra of Ce~3d levels. We present results obtained for the 3d$_{5/2}$ component. Spin-orbit splitting is equal to 18.5~eV.}
\begin{center}
\begin{tabular}{l l l l}
\hline
\hline
3d$_{5/2}$ & $f^0$ &$f^1$&$f^2$ \\
\hline
\hline
$E_B$ [eV]&897.45&884.50&881.97\\
$I$ [a.u.]&0.06&1&0.37\\
\hline
\hline
\end{tabular}
\end{center}
\label{table:xps}
\end{tablehere}
Satellites corresponding to the~f$^0$ and~f$^2$ final states are not fully resolved --- they are visible as shoulders of the intense f$^1$ line.  According to the theoretical model~\cite{fuggle83, gunnarsson83}, the intensity ratio $r_1=I(f^0)/[I(f^0)+I(f^1)+I(f^2)]$ yields information about the occupation of Ce~4f shell $n_f=1-r_1$. We have obtained the value~$n_f$=0.96. The hybridization energy can be estimated from the ratio $r_2=I(f^2)/[I(f^1)+I(f^2)]$. The value of~$r_2$ equal to~0.27 corresponds to the hybridization energy equal to~139~meV. 
	
	%fig4
\begin{figurehere}
\centering\includegraphics[width=0.9\linewidth]{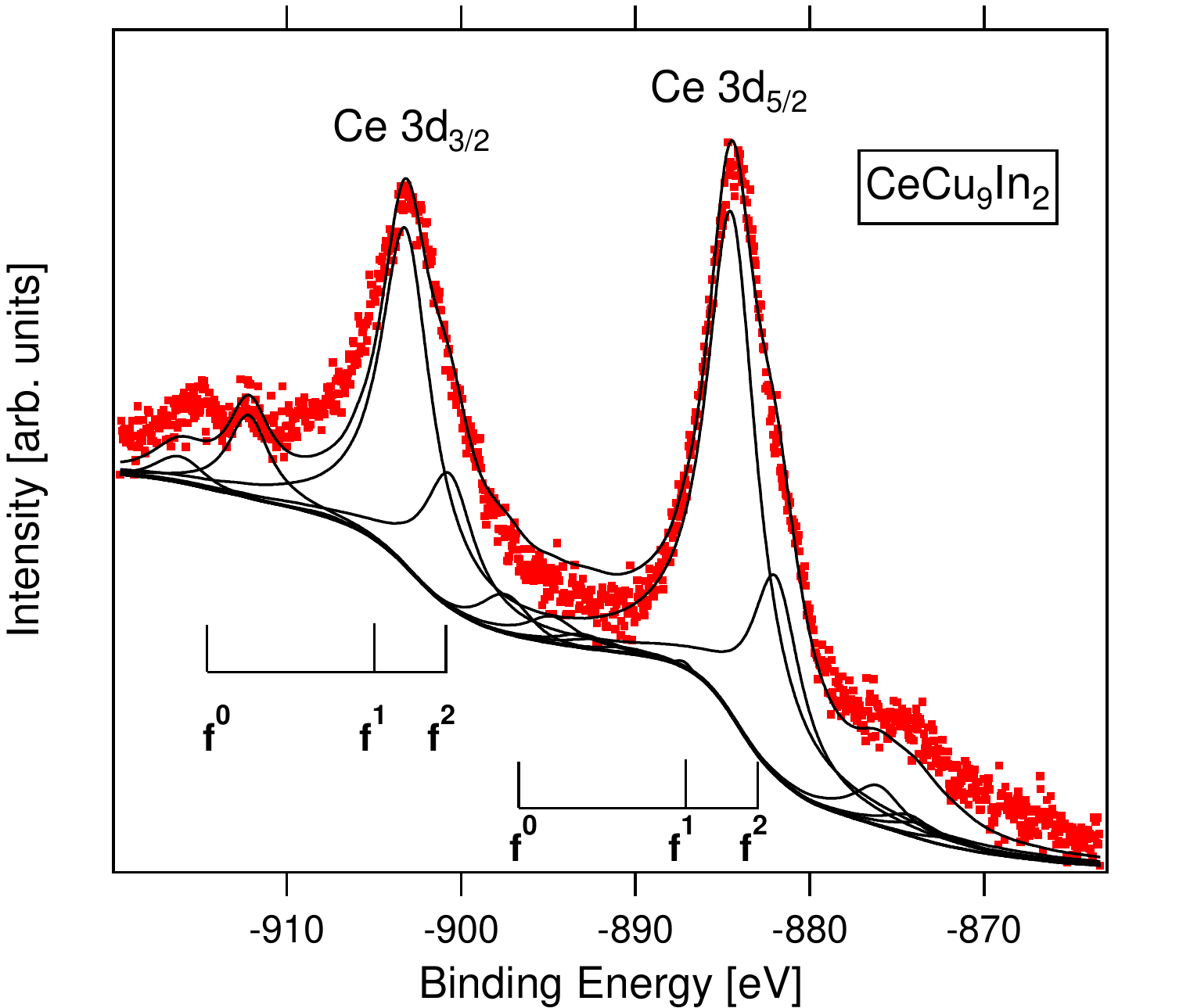}
\caption{(Color on-line) XPS spectra of the Ce~3d levels in CeCu$_9$In$_2$ obtained with Mg~K$_{\alpha}$ radiation at the temperature equal to~12.5~K. Experimental data are denoted by red dots, while solid black lines correspond to the fitted model.}
\label{fig:XPS}
\end{figurehere}

\subsection{Theoretical calculations}
\begin{figure*}
\centering\includegraphics[width=0.85\linewidth]{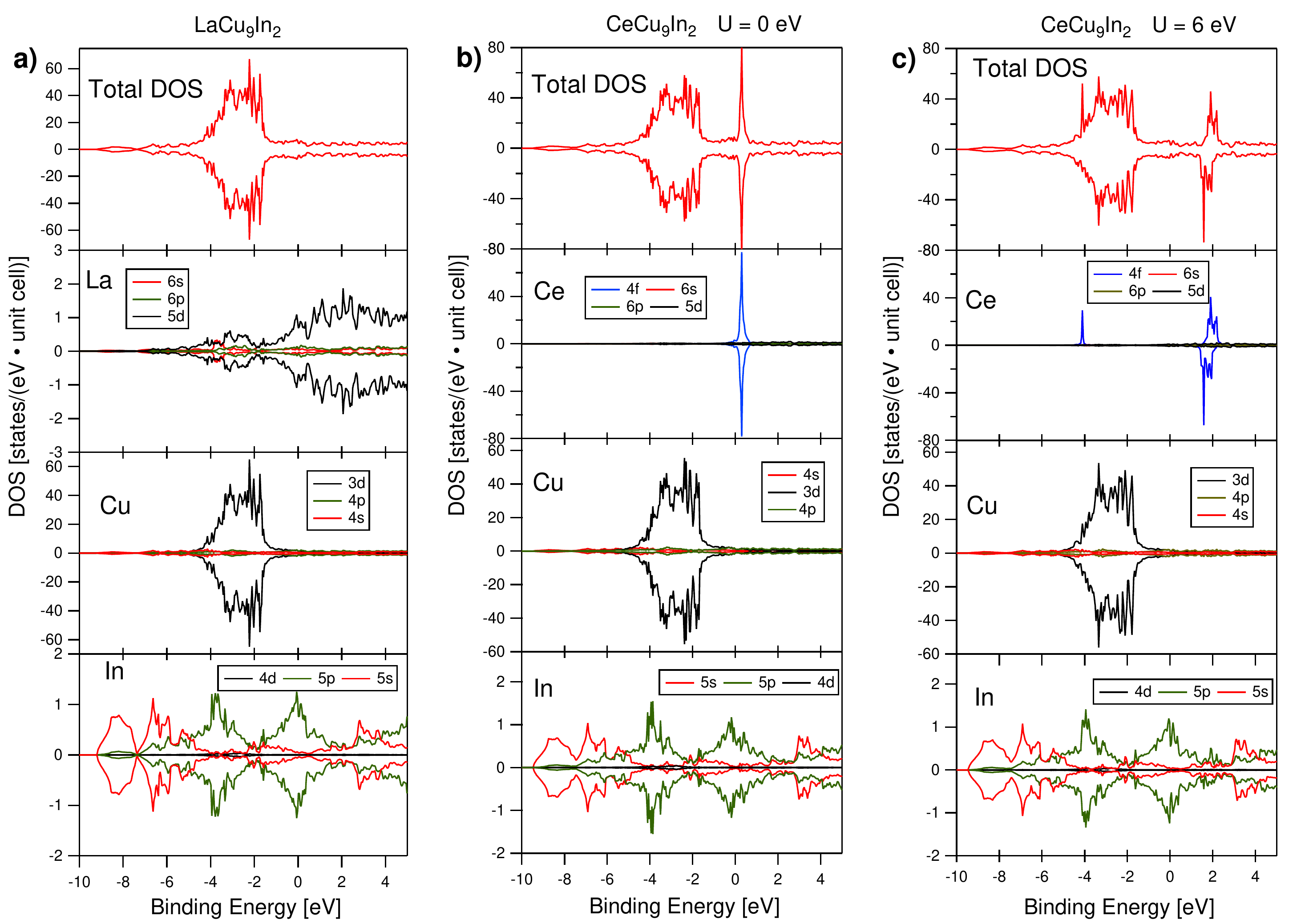}
\caption{(Color on-line) Total and partial densities of states calculated with application of FPLO code for (a) LaCu$_9$In$_2$ and (b) CeCu$_9$In$_2$ without~($U=$0 eV) and~(c) with~($U=$6 eV) additional correlations. Opposite spin directions are marked by positive and negative values of density of states.}
\label{fig:DOS}
\end{figure*}

	Theoretical calculations have been performed in order to get insight into the electronic structure of the studied compounds and to have a comparison with our experimental results. Total and partial densities of states, calculated for LaCu$_9$In$_2$ and CeCu$_9$In$_2$ with application of FPLO code, are shown in~\hyperref[fig:DOS]{Fig.~\ref*{fig:DOS}}. In case of CeCu$_9$In$_2$ densities of states have been calculated within two computational schemes: without interaction between f~electrons (U=0~eV, \hyperref[fig:DOS]{Fig.~\ref*{fig:DOS}~b}) and with Coulomb repulsion on 4f~shell (U=2~eV – not shown, U=6~eV, \hyperref[fig:DOS]{Fig.~\ref*{fig:DOS}~c}). 

	Density of states in the valence band of LaCu$_9$In$_2$ is dominated by the Cu~3d states~(\hyperref[fig:DOS]{Fig.~\ref*{fig:DOS}~a}). There is some humped structure build up from those states, mostly located between~$-5$~eV and~$-1$~eV. In the vicinity of the Fermi level, La~5d, In~5s and 5p~states contribute to the total density of states. La~5d states extend mostly above the Fermi level, while In~5s and 5p are rather smeared over the whole valence band. 

	In case of CeCu$_9$In$_2$ we have performed calculations for different values of Coulomb repulsion~(U) between electrons on 4f~shell. Densities of states calculated for U=0~eV, are symmetric and one can notice that for Cu and In, they are roughly the same as in case of the compound with~La. In partial density originating from Ce, contribution from 4f~orbital character is very significant. Namely, there is a sharp peak, located at about 0.3~eV above the Fermi level. For~U=6~eV the density of states is considerably modified. While the partial densities of states related to Cu and In remain roughly unchanged, there is an important change in case of partial Ce~4f~DOS. This quantity is not symmetric with respect to the spin degrees of freedom. One can see that, for arbitrary chosen, ''up''~component of the spin, 4f~DOS has two dominating structures. The first one is the peak at binding energy equal to~$-4$~eV, the second one is the peaked structure, which is located above the Fermi level. The maximum of the latter structure is at about~2~eV. 4f~DOS related to ''down'' component of the spin has got only one, hump-like structure. This feature lies above the Fermi level, with the maximum at about~1.7~eV. The asymmetry of the~4f, as well as total DOS, can lead to a presence of some spontaneous magnetic moment in the ground state of the CeCu$_9$In$_2$ compound. We have also performed calculation for~U=2~eV, however they are not qualitatively different from those for~U=0~eV.

%fig5

%fig6
\begin{figure*}
\centering\includegraphics[width=0.8\linewidth]{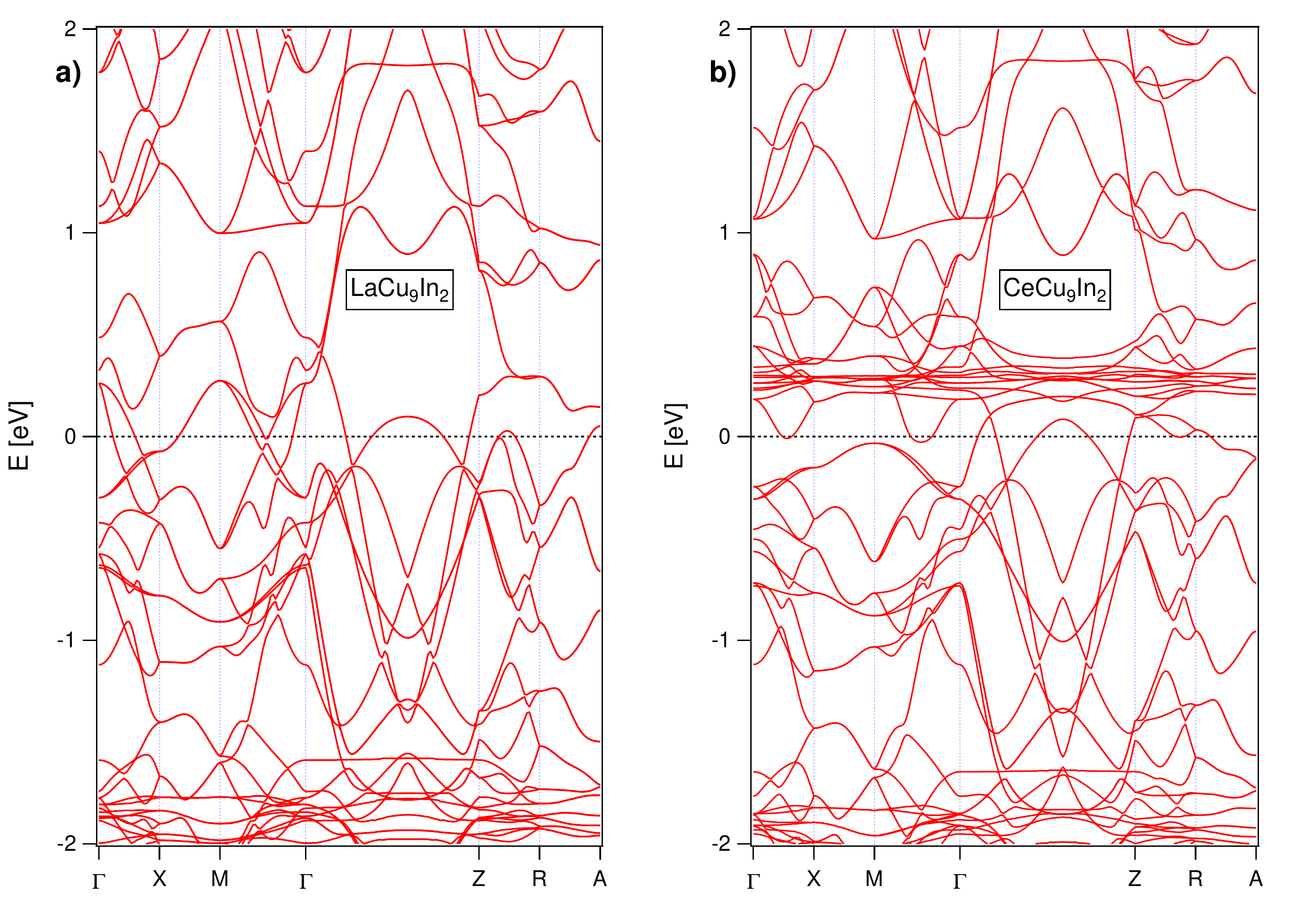}
\caption{(Color on-line) Ground state band structures of (a)~LaCu$_9$In$_2$ and (b)~CeCu$_9$In$_2$, calculated with application of Elk code. Non-magnetic ground state has been assumed.}
\label{fig:bands}
\end{figure*}
%\vspace{0.0cm}

\begin{figure*}
\centering\includegraphics[width=0.8\linewidth]{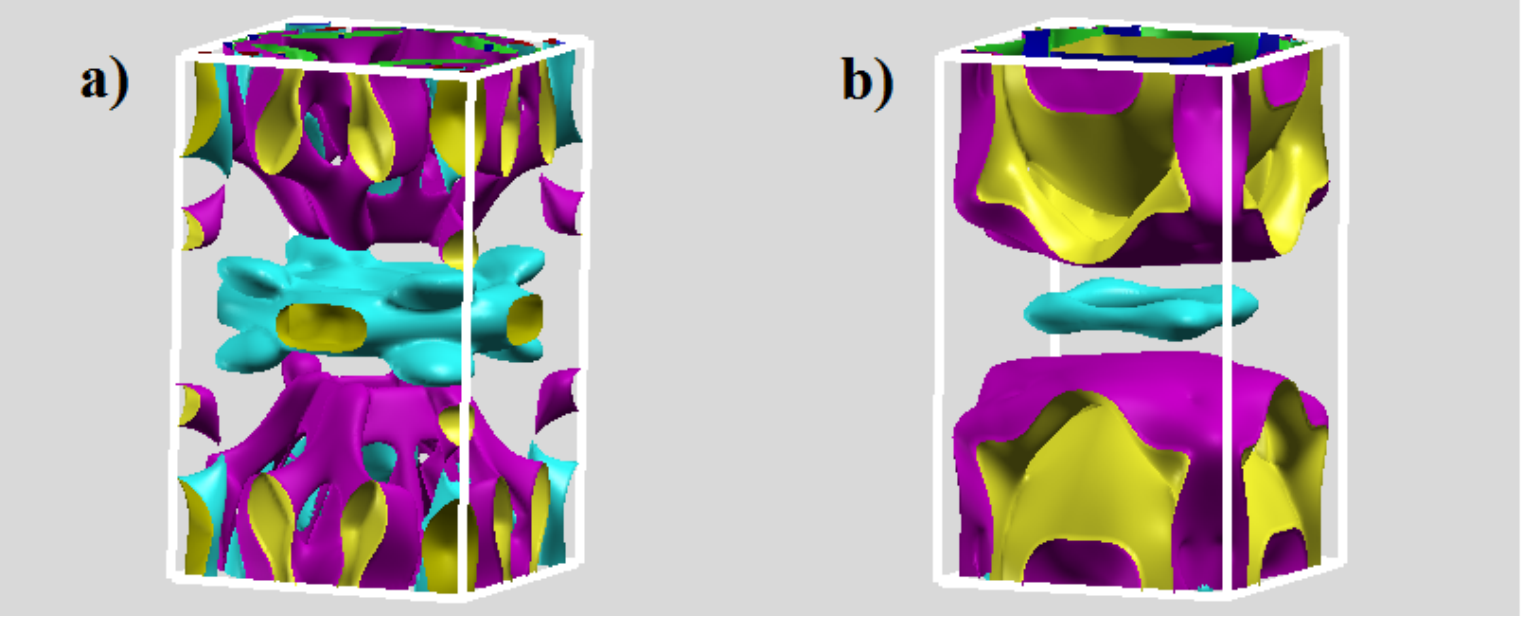}
\caption{(Color on-line) Fermi surface of (a)~LaCu$_9$In$_2$ and (b)~CeCu$_9$In$_2$, calculated with application of Elk code. Non-magnetic ground state has been assumed.}
\label{fig:FS}
\end{figure*}

	The band dispersion along some high symmetry lines inside first Brillouin zone (BZ) (\hyperref[fig:bands]{Fig.~\ref*{fig:bands}}), as well as FS (\hyperref[fig:FS]{Fig.~\ref*{fig:FS}}), have been calculated for CeCu$_9$In$_2$ and LaCu$_9$In$_2$. For both compounds, calculations have been done under assumption of non-magnetic ground state. One can notice, that far from the Fermi level, the bands have got similar dependency on the~\textbf{k}-vector. In the band structure of CeCu$_9$In$_2$, one can observe some structures which are absent in case of LaCu$_9$In$_2$. These are six flat bands visible above the Fermi level. They extend from about 0.2~eV to about~0.4~eV above the Fermi level. These states are mostly related to~4f~orbital character. So, according to the theoretical calculations, 4f~states in CeCu$_9$In$_2$ are not localized and are characterized by large effective masses. For both compounds one can observe some broad hole pocket around the center of the $\Gamma$-Z line. Beyond this fact, the band structure in the vicinity of the Fermi level is quite different for studied compounds and this fact is reflected in the calculated FS. FS (\hyperref[fig:DOS]{Fig.~\ref*{fig:FS}}) of both compounds is strongly anisotropic. In case of LaCu$_9$In$_2$ 4~bands contribute to FS, while in case of CeCu$_9$In$_2$ there are only 3~bands which cross the Fermi level. For the compound with~La, one can see a feature in the center of~BZ (around the $\Gamma$ point) – flattened in the direction perpendicular to~\textbf{c}$^{\star}$. This structure is surrounded, symmetrically in the \textbf{c}$^{\star}$ direction, by highly corrugated leaf-like structures. In case of CeCu$_9$In$_2$, there is a structure in the center of ~BZ, which resembles a doughnut which is flattened also in the direction perpendicular to \textbf{c}$^{\star}$. Moreover, one can see some almost flat parallel sheets of~FS, which are perpendicular to~\textbf{c}$^{\star}$.  This fact can be interpreted as an existence of~FS nesting in CeCu$_9$In$_2$.

%fig7

We have computed the Sommerfeld coefficient~($\gamma_{theor}$) with application of formula derived on the basis of free electron theory using calculated total densities of states at the Fermi level (D(E$_F$))  according to the equation~\cite{kittel96}:
\begin{equation}\gamma_{theor}=\frac{k_B^2 \pi^2}{3}   D(E_F) \end{equation}
The obtained values of the~$\gamma_{theor}$ coefficient are collated in the~\hyperref[table:gammaC]{Table~\ref*{table:gammaC}}. 	

%tutaj leci tabelka
%gamma
\begin{table*}
\centering
\caption{Comparison between experimental ($\gamma_{exp}$) and theoretical ($\gamma_{theor}$) values of the Sommerfeld coefficient for LaCu$_9$In$_2$ and CeCu$_9$In$_2$.}
\label{table:gammaC}
\begin{center}
\begin{tabular}{l c c c}
\hline
\hline
&U [eV]&$\gamma_{theor}$ [mJ/(mole$\cdot$K$^2$)]&$\gamma_{exp}$ [mJ/(mole$\cdot$K$^2$)]\\
\hline
\hline
LaCu$_9$In$_2$&0&14.08&14.0(1)\\
\hline
&0&19.76&\\
CeCu$_9$In$_2$&2&19.69&132(4)\\
&6&13.61\\
\hline
\hline
\end{tabular}
\end{center}
\end{table*}

One can notice, that in case of LaCu$_9$In$_2$, the value of~$\gamma_{theor}$ is in perfect agreement with the experimental one. This means that electrons in this compound are not strongly correlated and can be properly described with application of DFT based methods. On the contrary, in case of CeCu$_9$In$_2$, there is a large discrepancy between~$\gamma_{theor}$ and the experimental one. Even the introduction of correlation between 4f~electrons in the mean-field level does not lead to improvement of agreement between theory and experiment. It rather leads to greater discrepancy, than in case with~U=0. It can be explained as follows. For U=0~eV, DOS is symmetric with respect to the spin, and one can observe some peak-like structure in Ce~4f partial DOS, which has got a maximum above the Fermi level. However, there is some non-zero density of states at the Fermi level, because of the finite width of this structure. When Coulomb repulsion on the f shell is slightly higher than~0, then this peak moves above the Fermi level (calculations for U=2~eV are not shown), which corresponds to decrease of DOS at E$_F$ and finally to smaller value of~$\gamma_{theor}$, while the actual value of~$\gamma$ is enhanced by the correlation effects. Moreover, increase of U leads to splitting of the peak in vicinity of the Fermi level into two peaks because lower and higher Hubbard bands are formed. The first one is located below~E$_F$, while the second one  is placed above the Fermi level. Similar situation was observed previously in case of the CeNi$_9$In$_2$ compound~\cite{kurleto2015}. Discrepancy between theoretical and experimental value of Sommerfeld coefficient of CeCu$_9$In$_2$ is not startling for us and it testifies presence of strong correlations in this compound. 

\section{Conclusions}
In order to characterize the properties of the CeCu$_9$In$_2$ system we performed experimental studies of electrical resistivity, specific heat and electronic structure. The data are complemented by DFT~calculations. To separate an effect of 4f~electrons similar studies have been realized for the isostructural LaCu$_9$In$_2$ reference compound. The results indicate that CeCu$_9$In$_2$ is a Kondo lattice system with the coherence temperature~T$_{coh}$=45~K and Kondo temperature~T$_K\approx$5~K. Transition to the~AFM state is anticipated at~$T=$1.6~K. Properties of a Fermi liquid are observed at low temperatures. Lattice vibrations, Kondo effect, crystal field splitting and magnetism contribute to specific heat. Crystal field energy level scheme for CeCu$_9$In$_2$ is of the doublet-quartet type with splitting energy~$\Delta=$50.2(1.1)~K. Low temperature linear dependence of specific heat yields relatively high Sommerfeld coefficient of~132(4)~mJ/(mole$\cdot$K$^2$) what points to the enhancement of the effective mass of charge carriers. Another effects of coupling between 4f~electrons and conduction band are visible in the 3d~core level XPS spectra and in UPS spectra of valence band. The Kondo~peak, which is assigned usually to the~4f$^1$$_{5/2}$ final state is not visible in raw UPS data. However, increased spectral intensity near~E$_F$ is found in the extracted spectral contribution from 4f~electrons. We have observed peak at binding energy equal to~$-0.25$~eV, which is related to the~4f$^1$$_{7/2}$ final state. Theoretical calculations predict nesting of FS in case of CeCu$_9$In$_2$.

\section*{Acknowledgement}
This work has been supported by the National~Science~Centre,~Poland within the Grant no.~2016/23/N/ST3/02012. Support of the Polish Ministry of Science and Higher Education under the grant 7150/E-338/M/2018 is acknowledged. We are grateful to M.~Rams for measurements of specific heat at low temperature (0.4--10~K). The research was carried out with the equipment purchased thanks to the financial support of the European Regional Development Fund in the framework of the Polish Innovation Economy Operational Program (contract no.~POIG.02.01.00-12-023/08).

\bibliography{mybibfile}

\end{document}